# Optimization of the plasmonic properties of titanium nitride films sputtered at room temperature through microstructure and thickness control

Mateusz Nieborek[1], Cezariusz Jastrzębski[2], Tomasz Płociński[3], Piotr Wróbel[4], Aleksandra Seweryn[5], and Jarosław Judek[1,*]

[1] Institute of Microelectronics and Optoelectronics, Warsaw University of Technology, Koszykowa 75, 00-662 Warsaw, Poland.

[2] Faculty of Physics, Warsaw University of Technology, Koszykowa 75, 00-662 Warsaw, Poland.

[3] Faculty of Materials Science and Engineering, Warsaw University of Technology, Wołoska 141, 02-507, Warsaw, Poland.

[4] Faculty of Physics, University of Warsaw, Pasteura 5, 02-093 Warsaw, Poland.

[5] Institute of Physics, Polish Academy of Sciences, Aleja Lotników 32/46, 02-668 Warsaw, Poland.

[*] Corresponding author, e-mail address: jaroslaw.judek@pw.edu.pl.

## Abstract

A current approach to depositing highly plasmonic titanium nitride films using the magnetron sputtering technique assumes that the process is performed at temperatures high enough to ensure the atoms have sufficient diffusivities to form dense and highly crystalline films. In this work, we demonstrate that the plasmonic properties of TiN films can be efficiently tuned even without intentional substrate heating by influencing the details of the deposition process and entailed films' stoichiometry and microstructure. We also discuss the dependence of the deposition time/films' thickness on the optical properties, which is another degree of freedom in controlling the optical response of the refractory metal nitride films. The proposed strategy allows for robust and cost-effective production of large-scale substrates with good plasmonic properties in a CMOS technology-compatible process that can be further processed, e.g., structurized. All reported films are characterized by the maximal values of the plasmonic Figure of Merit (FoM = - $\varepsilon_1/\varepsilon_2$) ranging from 0.8 to 2.6, and the sample with the best plasmonic properties is characterized by FoM at 700 nm and 1550 nm that is equal 2.1 in both cases. These are outstanding results, considering the films' polycrystallinity and deposition at room temperature onto a non-matched substrate.

## Introduction

Titanium nitride belongs to an interesting class of plasmonic materials working in the ultraviolet, visible, and infrared spectral ranges and called refractory metal nitrides [1], [2], [3]. Despite generally worse plasmonic properties (a less negative real part $\varepsilon_1$ and a more positive imaginary part $\varepsilon_2$ of the dielectric function) than typically used noble metals like gold or silver, TiN poses some unique attributes that still make it attractive for photonic applications [4]. The most important feature is the compatibility with silicon technology, understood both as compatibility of the materials and fabrication techniques. Consequently, photonic structures or devices containing TiN as the plasmonic component could be potentially fabricated within existing industrial production lines, making the implementation straightforward and economically advantageous. The resistance to elevated temperatures, far higher than gold's and silver's melting temperatures, is another interesting and unique feature of this refractory metal nitride [5]. Decomposition temperatures reaching 2900 °C and the possibility of preservation of the optical properties up to even 1000 °C [6], [7], [8], [9] pave the way for applications requiring high operating temperatures [10], [11], [12]. Additionally, when considering the high abundance of constituent elements - titanium and nitrogen, in Earth's crust, cheap manufacturing methods, non-toxicity, and biological compatibility - titanium nitride appears to be a very economically and environmentally attractive plasmonic material.

Successful depositions of titanium nitride films with plasmonic properties have been widely reported in recent years. Among deposition techniques, one can include atomic layer deposition [13], [14], [15], [16], pulsed laser deposition [17], [18], molecular beam epitaxy [19], [20], and magnetron sputtering [21], [22], [23], [24], [25], [26], [27]. The deposition was performed onto a broad class of substrates, including rigid and flexible substrates [28], substrates that are crystallographically matched (e.g., MgO [29]), unmatched (e.g., silicon [30]), or intentionally highly mismatched [31], and even on polymers like PMMA [24]. The range of the substrate temperatures during depositions starts from room temperature to as high values as 1000 °C [28] – 1100 °C [31]. Typically, high-temperature epitaxy on a crystallographically matched substrate is expected to provide films with the best plasmonic properties. This fact might be, however, somehow problematic since high-temperature processes are not always possible or economically justified. For example, high-temperature processing is incompatible with complementary metal-oxide-semiconductor fabrication. In such cases, low-temperature deposition, particularly deposition at room temperature, might be the only reasonable choice. As regards the magnetron sputtering of plasmonic titanium nitride at room temperature, the following variations of the deposition were reported: 1) DC magnetron sputtering on Si(100) [21], 2) DC magnetron sputtering on MgO(100), c-plane sapphire, and Si(100) [22], 3) the high-power impulse magnetron sputtering onto B270 glass [23], 4) RF bias-free sputtering onto Si, quartz, $HfO_2$, and PMMA [24], 5) pulsed-DC sputtering with additional RF substrate biasing on soda-lime glass and Si(111) [25], and 6) reactive high-power impulse magnetron sputtering (R-HiPIMS) with glancing angle deposition (GLAD) on silicon (Si) wafer (100) [26].

In this work, we show a strategy for optimizing the plasmonic properties of relatively thick/nontransparent polycrystalline titanium nitride films deposited using the pulsed-DC reactive magnetron sputtering technique at room temperature on a non-matched substrate, i.e., Si(100) wafer with a 10 nm thick $TiO_x$ interlayer. Application of the dielectric buffer layer decreases the films' stress, allowing for successful TiN deposition on 4" Si wafers. The main advancement we report is the demonstration that the plasmonic properties of titanium nitride films can be efficiently tuned even without intentional heating by influencing the details of the deposition process and entailed films' stoichiometry and microstructure by controlling the working pressure and gas flows. We also discuss the dependence of the deposition time/films' thickness on the optical properties, which is another degree of freedom in controlling the optical response of the titanium nitride films [32, 33, 34].

Our idea/strategy can be understood within the framework of modern structure zone models (SZM), which try to relate films' microstructures and the parameters of the physical vapor deposition processes. Historically, the idea of SZM was first introduced by Movchan and Demchishin, who noticed that the critical parameter determining the films' microstructures is the ratio of the substrate temperature during the deposition process to the material's melting temperature. The so-called homologous temperature translates into the adatoms' surface mobility/diffusivity, directly determining what microstructure will be formed. In their work, Movchan and Demchishin identified/distinguished three types of possible microstructure that can be formed depending on the homologous temperature value [35]. Next, Thornton extended the SZM by noticing the importance of working gas pressure [36], whereas Mahieu raised a question concerning the energy and atomic fluxes towards the substrate [37, 38]. Finally, Anders generalized the SZM by suggesting that the homologous temperature should be replaced by "a generalized temperature, which includes the homologous temperature plus a temperature shift caused by the potential energy of particles arriving on the surface" and the pressure by "a normalized energy, describing displacement and heating effects caused by the kinetic energy of bombarding particles" [39]. The last sentence seems especially important since the kinetic energy distribution within the particle flux toward the substrate might be the missing link to explain why and how pressure and argon flow influence the film microstructure. We would like to make a digression here. The relation between the pressure and the distribution of the kinetic energy of the particle flux toward a substrate was recently examined in a completely different context. Johansson [40] and Pliatsikas [41] investigated sputter-induced damage to the substrate's surface during the deposition using graphene as the damage indicator. Both publications conclude that the kinetic energy distribution within the Ar ions depends directly on the pressure. At low working pressures, the ballistic damage due to Ar bombardment is the highest due to the increased probability of highly energetic Ar ions within the plasma.

For the sake of order, we note that the complementary study on the influence of substrate temperature on the microstructure and optical properties of titanium nitride films was already reported [42].

# Results and discussion

*microstructure and plasmonic properties of titanium nitride films as a function of $N_2$ and Ar flow values*

In the first part of our experiment, we analyze how parameters of the employed physical vapor deposition process, i.e., argon flow, which determines the working pressure, simultaneously with nitrogen flow, influence the stoichiometry, microstructure, and optical properties of the deposited films and if any correlation between these three properties can be noticed. We note that all samples whose properties are reported in this subchapter were deposited with constant time, and in consequence, their thickness varies.

Figure 1 presents data that can be helpful for qualitative analysis. Figure 1a-d shows pictures of four series of samples deposited for four different Ar flow values: 10 sccm, 25 sccm, 50 sccm, and 100 sccm. Within each series, samples are deposited with such nitrogen flow values to get stoichiometric and non-stoichiometric samples, both with nitrogen deficiency and excess. It can be perceived that samples deposited for the lowest nitrogen flow are silver in color, samples deposited for the moderate nitrogen flow have golden luster, whereas samples deposited for the highest nitrogen flow are brownish. It is because changing nitrogen flow for the selected value of the argon flow allows for the stoichiometry control, and stoichiometry-dependent changes in films' color is a typical and widely observed behavior for refractory metal nitrides due to the composition-dependent shift in screened plasma frequency value [43]. Figures 1e-h show SEM images of samples' cross-sections. A vertical columnar structure can be distinguished for all samples, but drawing more precise conclusions about the microstructure is difficult. The exception is the surface topography characterized by faceting, which depends on Ar flow and is most pronounced for the 100 sccm Ar flow value. The increase in surface faceting as the Ar flow increases is rendered in AFM images shown in Figure 1i-l, also as the increase in surface roughness. To study the microstructure, we performed the XRD measurements, the results of which are shown in Figures 1m-p as XRD patterns. The changes in relative (111), (200), and (220) peak intensities related to the changes in crystallographic orientations of the nanocrystallites within the polycrystalline film as well as the peak broadening as the Ar flow decreases, can be easily noticed. Particularly, one can distinguish a transition from one narrow (111) peak acquired from samples deposited with 100 sccm Ar flow to three broad (111), (200), and (220) peaks acquired for samples deposited with 10 sccm Ar flow.

To analyze all these data quantitively, we plotted the deposition rate value estimated from the SEM images, RMS values from AFM images, and nanocrystallite size calculated with the Scherrer equation and relative intensities of the XRD peaks in Figure 2. Figure 2a reveals a strong dependence of the deposition rate on nitrogen flow for all Ar flow values suggesting that the deposition process of the titanium nitride films in all cases occurs on the border between the metallic and dielectric mode of the sputtering process – in the transient mode. The range of changes in the deposition rate when changing nitrogen flow is qualitatively the same regardless of the Ar flow value. What is changing is only the nitrogen flow value that corresponds to the deposition of the stoichiometric films. This high sensitivity of the deposition rate on nitrogen flow for the selected value of the Ar flow can be explained by the increasing coverage of the titanium target by titanium nitride, known as target poisoning. The high variation in the deposition rate is, however, not very strongly rendered in the surface roughness value depicted in Figure 2b. In the case of surface roughness, the influence of the nitrogen flow is minor compared to the strong influence of the Ar flow. The XRD results shed more light on this behavior. Figures 2c-f illustrate two simultaneous processes that occur when decreasing the Ar flow (the working pressure): the decrease in the nanocrystallite size and the increase in the fraction of other than (111) nanocrystallite orientations. For the highest Ar flow (the highest working pressure of about 0.36 Pa), all nanocrystallites are characterized by an average size of 37 nm and are oriented along (111) orientation despite the samples' stoichiometry. Decreasing the Ar flow to 50 sccm, which results in a decrease in working pressure to 0.15 Pa, leads to a decrease in the average nanocrystallite size to 22 nm and the emergence of a second (200) nanocrystallite orientation. Further decrease of the Ar flow to 25 sccm resulting in working pressure of 0.04 Pa leads to an average nanocrystallite size of 18 nm and further decrease of the (111) nanocrystallites orientation fraction. The lowest Ar flow of 10 sccm results in a working pressure value below 0.013 Pa, nanocrystallite size as small as 8 nm, and the emergence of a distinct (220) peak.

Our interpretation of all these observations is that working pressure control through Ar flow control allows us to change the kinetic energy distribution of the elements constituting the plasma and, consequently, the kinetic energy distribution in the particle flux approaching the substrate, according to modern structure zone models [36-39]. The lower the working pressure, the higher the average kinetic energy of atoms/ions arriving at the film's surface. This kinetic energy can be an additional energy source for the adatoms improving their surface mobility, which should lead to improved crystalline quality of the deposited films. But simultaneously, the high-energy bombardment of the surface may deteriorate its crystalline structure. The extreme case is surface etching [39] due to the bombardment of the surface by high-energy ions/atoms, leading to consequent adatoms removal/ejection. In our case, decreasing the working pressure through the Ar flow value decrease seems not to lead to surface etching, which would manifest as a decrease in the deposition rate. On the contrary, the deposition rate for the lowest Ar flow seems to be higher than for other Ar flow values. Thus, we are not in the etching regime. On the other hand, it seems that the crystalline quality of the films increases neither when decreasing the working pressure since the average nanocrystallite size decreases. We are neither convinced if the emergence of the (200) nanocrystallite orientation, which previously was reported as the positive effect of the higher energy ion bombardment [44] (the (200) surface orientation is characterized by lower surface energy than the (111) one), is a sign of increased crystal quality. We are rather inclined to conclude that we are observing progressive amorphization of the deposited films, understood as the decrease in the average nanocrystallite size and the emergence of random nanocrystallite orientation due to decreased working pressure through the Ar flow value decrease, which results in an increased surface bombardment by high energy ions/atoms. In other words, by changing the working pressure by the Ar flow, we can control the amorphization state of the deposited titanium nitride films by varying the kinetic energy distribution in the particle flux approaching and impacting the substrate. The lower the Ar flow, the lower the pressure and the more highly energetic ions/atoms impact the surface, and thus more amorphous/less crystalline films are deposited.

Optical properties being a result of the ellipsometric measurements are shown in Figure 3 as the real $\varepsilon_1$ and imaginary $\varepsilon_2$ part of the dielectric function, and the plasmonic Figure of Merit defined as the ratio of the minus real to the imaginary part of the dielectric function $-\varepsilon_1/\varepsilon_2$. All presented traces are dominated by the strong Drude component – a characteristic feature of the metallic titanium nitride with decent plasmonic properties. The free electron contribution manifests as a monotonically decreasing negative value of the $\varepsilon_1$ and a monotonically increasing positive value of $\varepsilon_2$. For plasmonic application, the lower (more negative) value of $\varepsilon_1$, the better; similarly, the lower (less positive) value of $\varepsilon_2$, the better. The plasmonic figure of merit defined above is a value related to the localized surface plasmon resonance phenomena and it is one of the most popular choices for comparing plasmonic materials. However, choosing other, more application-specific Figures of Merit is also possible [45]. To compare the optical properties more quantitatively, we plotted selected parameters calculated from the dielectric function in Figure 4.

Figure 4a illustrates the wavelength at which the real part of the dielectric function equals zero as a function of nitrogen flow for selected values of Ar flow. Since there is a strong correlation between the λ@$\varepsilon_1$=0 and the stoichiometry [43], deposited titanium nitride films are stoichiometric (480 nm) and non-stoichiometric, both with the nitrogen excess (λ > 480 nm), and the deficiency (λ < 480 nm), for every Ar flow value. Samples deposited with the lowest nitrogen flow value within selected Ar flow have nitrogen deficiency, whereas samples deposited with the highest nitrogen flow value within selected Ar flow have nitrogen excess. Figure 4b illustrates the value of the imaginary part of the dielectric function at the same wavelength at which the real part equals zero. In our case, when the deposition process is performed with constant time, the lowest values are achieved for the most stoichiometric samples. The plasmonic figure of merit is shown in Figure 4c. As can be seen, the FoM takes the highest values for the most stoichiometric samples within each of the Ar series, but between the series, the FoM for samples deposited with Ar flow that equals 25 sccm takes the highest values. These are very interesting results since, whereas the fact that stoichiometric samples have the best plasmonic properties is somehow intuitive, the existence of the optimum value of the Ar flow that controls the working pressure, for which plasmonic properties are the best, is surprising. Similarly surprising is that the samples deposited with 100 sccm Ar, which are the most crystalline, have the worst plasmonic properties. And the improvement of the plasmonic properties correlates with the progressive amorphization of the films due to the decrease of the working pressure, at least to some critical point (Ar 25 sccm), after which the optical properties deteriorate (Ar 10 sccm). Similar conclusions can be drawn by analyzing the parameters of the Drude component of the dielectric function, which governs the optical properties of TiN in the infrared range: the plasma energy denoted as $E_{pu}$ and related to the carrier concentration and its effective mass, and the damping coefficient denoted as $\Gamma_D$ and related to various scattering mechanisms of free electrons. Formally, we are using the following equation:

$$\varepsilon_{\text{Drude}}(E) = \varepsilon_\infty + \frac{E_{pu}^2}{-E^2 - i\,E\,\Gamma_D}, \quad (1)$$

where $E$ stands for the photon energy, and $i$ stands for the imaginary unit. As can be seen in Figure 4d, the average value of the plasma energy increases when the Ar flow decreases from 100 sccm to 25 sccm, but when the Ar flow reaches 10 sccm, the $E_{pu}$ seems to decrease slightly, at least for the stoichiometric samples. Similarly behaves the Drude damping coefficient plotted in Figure 4e – its value decreases when the Ar flow decreases from 100 sccm to 25 sccm, but when the Ar flow reaches 10 sccm, the $\Gamma_D$ seems to increase its value slightly. So the low-pressure sputtering, which means increased surface bombardment by the high-energy particles, might improve the average metallicity of the TiN film, at least to some point. However, simultaneously progressive amorphization, which also means a decrease in the average nanocrystallite size as the working pressure decreases, may lead to breaking down this positive trend for the smallest nanocrystallites, as seen in the example of the data acquired for the 10 sccm Ar flow.

We also note that the last conclusion partially aligns with the literature that reports the decrease in the resistivity value as the working pressure during the deposition of the TiN samples decreases [44]. However, since the sputtering pressure in the cited work is decreased from 0.65 Pa to only 0.13 Pa, only the positive effect of the lower working pressure (decrease in the resistivity value) is observed. In our work, we are able to grasp the breaking down of the positive influence of lower working pressure because the minimal value of the pressure equals 0.013 Pa, one order of magnitude lower than in the cited work.

*evolution of the plasmonic properties of the stoichiometric titanium nitride films with thickness*

In the second part of our experiment, we analyze how the deposition time in the employed physical vapor deposition system, which translates directly into films' thickness, influences the optical properties of the stoichiometric films characterized by the different microstructures.

Figure 5 shows data helpful for qualitative analysis. The SEM images of the thickest samples (2.2 μm, 2.1 μm, 3.1 μm, and 3.3 μm) shown in Figure 5a-d reveal columnar structure without any significant changes during deposition, and the AFM images shown in Figures 5e-h illustrate the surface topography with clear evolution of the surface faceting. The real part of the dielectric function shown in Figure 5i-l seems to not depend significantly on the films' thickness but depends on the microstructure. The imaginary part of the dielectric function, shown in Figures 5m-p, on the other hand, depends both on the films' thickness and microstructure. Consequently, the plasmonic figure of merit depends on the microstructure and films' thickness. To analyze these data quantitively, we plotted selected parameters characterizing the dielectric function in Figure 6.

Figure 6a shows maximal values of the Figure of Merit for samples obtained for varying deposition time that directly translates into the films' thickness for all four series of Ar flow. As can be seen, the plasmonic properties improve with the films' thickness, particularly, exponential saturation of the plasmonic Figure of Merit versus the films' thickness can be observed (the solid line in Figure 6a is the fitted exponential function). Similarly to the first part of our experiment, samples characterized by the best plasmonic properties were deposited with 25 sccm Ar flow, and the plasmonic figure of merit for series deposited with lower (10 sccm) and higher (50 sccm, 100 sccm) Ar flow is worse. Thus, the shown data allow for upholding the thesis on the influence of the microstructure on plasmonic properties. Our result in part related to the thickness dependence is qualitatively similar to the result obtained for TiN film grown using the ALD method on a sapphire substrate at 450 °C and which was interpreted as the effect of lower quality of the layers near the substrate and the increase of the broadly understood quality as the thickness increases [32]. We note, however, that in the cited work, the thickness of the films equals up to 100 nm, whereas in our case, it is in the micrometer range; particularly for our samples, the saturation/stabilization of the plasmonic figure of merit occurs for films thicker than 1 μm.

Figure 6b illustrates changes in the RMS value as a function of thickness. A linear increase in the film's roughness is observed, and the surface roughness for the selected thickness correlates with the working pressure at which the samples were deposited – the lower the pressure, the lower the roughness. So, the RMS data correlate well with the nanocrystallite size calculated from the XRD spectra. The increase in the surface roughness as the thickness increases is a typical observation. The possible explanations include the accumulation of the effect of the structural defects and increased grain size.

To provide further insight into made observations, we analyze the plasma energy value and the damping coefficient from the Drude model shown in Figures 6c and d, respectively. The plasma energy value decreases with thickness for samples deposited with the 10 sccm, 25 sccm, and 100 sccm Ar flow, whereas the results for the 50 sccm Ar flow are ambiguous. The Drude damping coefficient exponentially decreases with thickness for all Ar flow values (the solid line in Figure 6d is the fitted exponential function). A decrease in plasma energy means a decrease in the averaged film metallicity, which seems to be a rather negative effect. In contrast, a decrease in the damping coefficient means a decrease in the scattering rate, which is a positive effect. Similar effects have been reported by Saha et al. [34] for the polycrystalline TiN films deposited on silicon at 800 °C with thicknesses up to 200 nm. The interpretation was that the increased grain size for thicker samples results in reduced collisions of charge carriers within the grains (the decrease in the damping constant), and the columnar growth with many grain boundaries results in reduced film density and facilitates the oxygen diffusion into the TiN film (the decrease in the plasma energy). This interpretation is also possible for our results since it is supported, e.g., by the surface roughness analysis. Still, other explanations, also those related to process and technology details, are not undoubtedly excluded [46]. Nevertheless, additional structural analysis using STEM microscopy has been performed for the thickest layer deposited with the 25 Ar flow to check if grain size indeed changes during the deposition process, as suspected.

As revealed in Figure 7a, the structure of the deposited polycrystalline layer consists of ultrafine elongated grains with a size that depends on the distance from the substrate (deposition time). The smallest grains can be found at the interface with the substrate, and as the layer becomes thicker, the grain size increases, reaching its maximum near the surface. Figures 7b-d show three high-resolution images taken near the interface, in the middle of the layer, and near the surface. These images were transformed using the Fast Fourier Transform (FFT) to get diffraction patterns, shown in Figures 7e-g. These patterns are very complex due to the high density of structural defects and low-angle grain boundaries, and thus, rigorous interpretation is impossible without numerical analysis. However, still some semi-qualitative analysis can be performed. At the interface with the substrate, the FFT diffraction pattern (Figure 7g) is characterized by continuous rings, which suggests a high number of small grains. Their thickness was estimated to be in the 8 - 12 nm range, and their length was estimated not to exceed 50 nm. These grains are elongated along the growth direction. In the middle of the sample, 50 nm thick and 200 nm long grains can be found, but we note that despite the increase in the typical grain size, these larger grains are still surrounded by smaller grains. The rings in Figure 7f became discontinuous, meaning the number of grains decreased in the scanning area significantly compared to the interface. Near the surface, the grain size increases even more, and the rings in Figure 7e are even more discontinuous when compared to Figure 7f. The individual bright spots in the FFT diffraction pattern can be attributed to 200 nm thick and 500 nm long grains that can be occasionally identified. But as in the middle of the sample, they are still surrounded by smaller grains. Our direct observations of the increase in the grain size are an expected result and are in line with previous literature reports. However, recalling the optical context of our work, it is the first demonstration of a correlation between the grain size evolution and changes in the plasmonic performance of the films.

*plasmonic gratings on the surface of titanium nitride thick film*

One of our work aims is to develop a silicon technology-compatible substrate with plasmonic properties competitive with gold and silver that can be further safely processed, e.g., structurized. Therefore, in this subchapter, we theoretically analyze the optical response (reflection vs. wavelength) of one-dimensional plasmonic grating [47] fabricated on the surface of the titanium nitride thick/nontransparent film. For simulations with Comsol Multiphysics, we used the dielectric function describing the thickest TiN film deposited with 25 sccm Ar flow. The Drude parameters equal $E_{pu} = 7.4$ eV and $\Gamma_D = 0.35$ eV, and the corresponding refractive index $n$ and extinction coefficient $k$ are shown in Figure 8a. The blue region indicates the spectral range for which $n < 1$, i.e., approximately from 520 nm up to 1020 nm. The simulated plasmonic grating is assumed to consist of an infinite one-dimensional array of rectangular cross-section infinite nanogrooves that are etched on the surface of the infinite (thick enough/nontransparent) TiN substrate with the following dimensions: the period equals 500 nm, the depth of the nanogroove equals 70 nm, and the width varies from 40 nm to 400 nm. Exemplary electric field distribution when the polarization of the incident light is parallel (TE) and perpendicular to the nanogrooves (TM) within one unit cell is shown in Figure 8b, whereas the simulated reflection values at normal incidence are shown in Figure 8c. The presented data clearly demonstrate surface plasmon resonance (SPR) excitation manifested by the minimum reflection occurring only for the TM polarization. The presence of the grating assures momentum matching of the surface plasmon-polaritons (SPPs) and the incident photons, and the efficiency of the photon-plasmon coupling for a given grating period can be optimized by the adjustment of a nanogroove width. For optimal conditions, the energy of the incident wave is totally transferred to the propagating surface mode, resulting in zero reflection at the resonant wavelength (approximately from 570 nm to 720 nm). These results confirm that the optimized material might be an excellent counterpart of the noble metals for fabricating plasmon-enhanced devices functional in the VIS-NIR spectral range.

Structurization of the developed material might also be used to fabricate the nanostructures allowing the excitation of localized surface plasmons (LSPs) exhibiting stronger light confinement at the metal-dielectric interface in comparison to the propagating SPP modes. For example, localized surface plasmon resonance (LSPR) with null reflection at a particular wavelength has been recently experimentally demonstrated for the HfN nanodisk array [48].

# Conclusions

We demonstrate an optimization strategy for depositing relatively thick titanium nitride film with competitive plasmonic properties at room temperature onto a non-matched substrate. We exploit the fact that proper control of the selected process parameters allows for controlling the films' stoichiometry, microstructure, and thickness – three properties that significantly influence the optical properties of polycrystalline titanium nitride films. Consequently, the sample with the best plasmonic properties is characterized by a maximum FoM value of 2.6 and FoM value of 2.1 at 700 nm and 1550 nm. Simultaneously, we made four interesting observations that shed some light on the physical phenomena occurring during the deposition process. The first two are related to the fact that working pressure control through Ar flow control determines the kinetic energy distribution in the particle flux approaching and impacting the substrate. Thus, as the working pressure decreases, we are observing the increase in the averaged metallicity of the films, understood as the increase in the Drude plasma energy value, and the progressive amorphization understood as the decrease in the average nanocrystallite size and the emergence of random nanocrystallite orientation. The last two observations are related to the changes in the film surface as the film thickness increases: a decrease in the averaged metallicity of the films, which seems to be a negative effect, and a decrease in the damping constant, which is a positive effect. Both originate from the directly observed increase in the grain size as the deposition time increases.

Experimental

*Deposition of the TiN films*

All titanium nitride films were deposited using a pulsed-DC reactive magnetron sputtering technique in the Oxford Plasmalab System 400 Sputter tool. The Advanced Energy Pinnacle Plus+ power supply (100 kHz pulsing frequency and 4 μs off-time) delivered the power to the chamber. The deposition was performed at room temperature; however, some non-essential and non-intentional sample heating was observed during the deposition. The sputtering was performed from a pure 8 inches diameter titanium target. Argon was the main sputtering gas, whereas adding gaseous nitrogen made the deposition process reactive. The chamber base pressure value was below $5.0\times10^{-7}$ Torr before and after every deposition. During the deposition process, the pressure was not controlled by the pressure controller but by the Ar flow. The exact values of the working pressure are below 0.1 mTorr (below 0.013 Pa) for 10 sccm Ar flow, 0.3 mTorr (0.04 Pa) for 25 sccm Ar flow, 1.1 mTorr (0.15 Pa) for 50 sccm Ar flow, and 2.7 mTorr (0.36 Pa) for 100 sccm Ar flow.

All reported in this paper titanium nitride films were deposited on a typical silicon wafer with (100) orientation. However, the crystalline titanium nitride films deposited in our system directly on the monocrystalline silicon substrate (even with the native oxide layer) are characterized by enormous stress, which typically leads to damage through cracking. Thus, a 10 nm thick TiO$_x$ interlayer was deposited on the silicon wafer before the proper titanium nitride film was deposited to mitigate this problem.

*Characterization*

Surface topography was investigated using Bruker Dimension Icon Atomic Force Microscope in the ScanAsyst mode. Scan areas are set to 2 μm × 2 μm (512 points × 512 points).

The X-Ray diffraction (XRD) measurements were performed with the PANalytical X'Pert Pro MRD diffractometer equipped with the radiation-generating tube of the wavelength of 1.54056 Å, a hybrid two-bounce Ge (220) monochromator and a PIXcel detector with the parallel plate collimator with the 0.4-rad Soller slits and a 0.18-deg divergence slit. All scans were performed in Bragg-Brentano geometry. The peak profiles in the XRD patterns were described by the Voigt function. The empirical Sherrer's formula was employed to estimate the size of the nanocrystallites. Since all the raw XRD patterns are dominated by the silicon (400) reflection from the (100) oriented substrate, the (400) reflection was numerically subtracted from all the XRD patterns to make the plots more readable. Next, the patterns were baseline-corrected and normalized to the titanium nitride (111) peak.

For the TEM observations, the FIB lamella was prepared by the lift-out technique perpendicular to the sample surface. The microstructure observations were performed by using the HR TEM/STEM microscope model Spectra 200 made by Thermo Fisher Scientific. The observations were carried out at 200kV in STEM mode using HAADF and BF detectors. The Fast Fourier Transform was obtained from high-resolution images.

Optical characterization was performed using Woollam RC2 spectroscopic ellipsometer. The $\Psi(\lambda)$ and $\Delta(\lambda)$ parameters were recorded in the range of 193 nm to 1690 nm for the following incident angles: 45°, 50°, 55°, 60°, 65°, 70°, and 75°. Modeling of the optical properties was performed in CompleteEASE software. Although most of the fabricated $TiN_x$ layers are expected to be opaque, some slight influence of the substrate on the extracted dielectric function was observed for the thinnest samples. Thus, in order not to introduce additional uncertainty related to the change of the optical model for different samples, our optical model takes into account all the constituents of the investigated structure, i.e., Si substrate, native $SiO_2$ layer, $TiO_x$ interlayer, the $TiN_x$ layer, and the roughness for all the samples. Dielectric constants of Si, $SiO_2$, and $TiO_x$ are taken from the Woollam database, while $TiN_x$ dielectric function is modeled with the use of the General Oscillator model where the real part of the permittivity is described by the $e_\infty$ that is permittivity at an infinite frequency and two poles that are equivalent to a Lorentz oscillator with zero broadening, that are placed outside of the measured spectral range. The imaginary part of permittivity in the UV-VIS is described by four Lorentz oscillators, and the metallic properties in the NIR range are parametrized by the Drude oscillator. The roughness is modeled as an effective layer assuming 50 % content of voids and underlying material, and its optical constants are calculated using the Bruggeman Effective Medium Approximation. It is worth mentioning that although the measured $TiN_x$ samples might be considered semi-infinite, the pseudo-dielectric function approximation is not applicable in our case. First, because our samples are nontransparent and highly absorbing, thus the whole ellipsometric information is acquired from the surface. Second, the surface roughness characterizing our samples is large enough to introduce differences between the results obtained using the pseudodielectric approximation and the different ellipsometric models that try to include the effect of finite surface roughness. Further discussion can be found in the Supplementary Information.

## Data availability

Data sets generated and/or analyzed during the current study are available from the corresponding author on reasonable request.

## Acknowledgment

This work was supported by the POB FOTECH-3 project entitled "Plasmons and polaritons on nanostructured surfaces of IVb metal nitrides" granted by the Warsaw University of Technology within "The Excellence Initiative – Research University" program.

## Contributions

**M. N.**: Investigation, Data Curation, Writing - Original Draft, Writing - Review & Editing; **C. J.**: Investigation, Writing - Review & Editing; **T. P.**: Investigation, Writing - Review & Editing; **P. W.**: Methodology, Formal analysis, Investigation, Writing - Original Draft, Writing - Review & Editing; **A. S.**: Investigation, Writing - Review & Editing; **J. J.**: Conceptualization, Methodology, Writing - Original Draft, Writing - Review & Editing, Supervision, Project administration, Funding acquisition.

## Competing Interest

The authors declare that they have no known competing financial interests or personal relationships that could have appeared to influence the work reported in this paper.

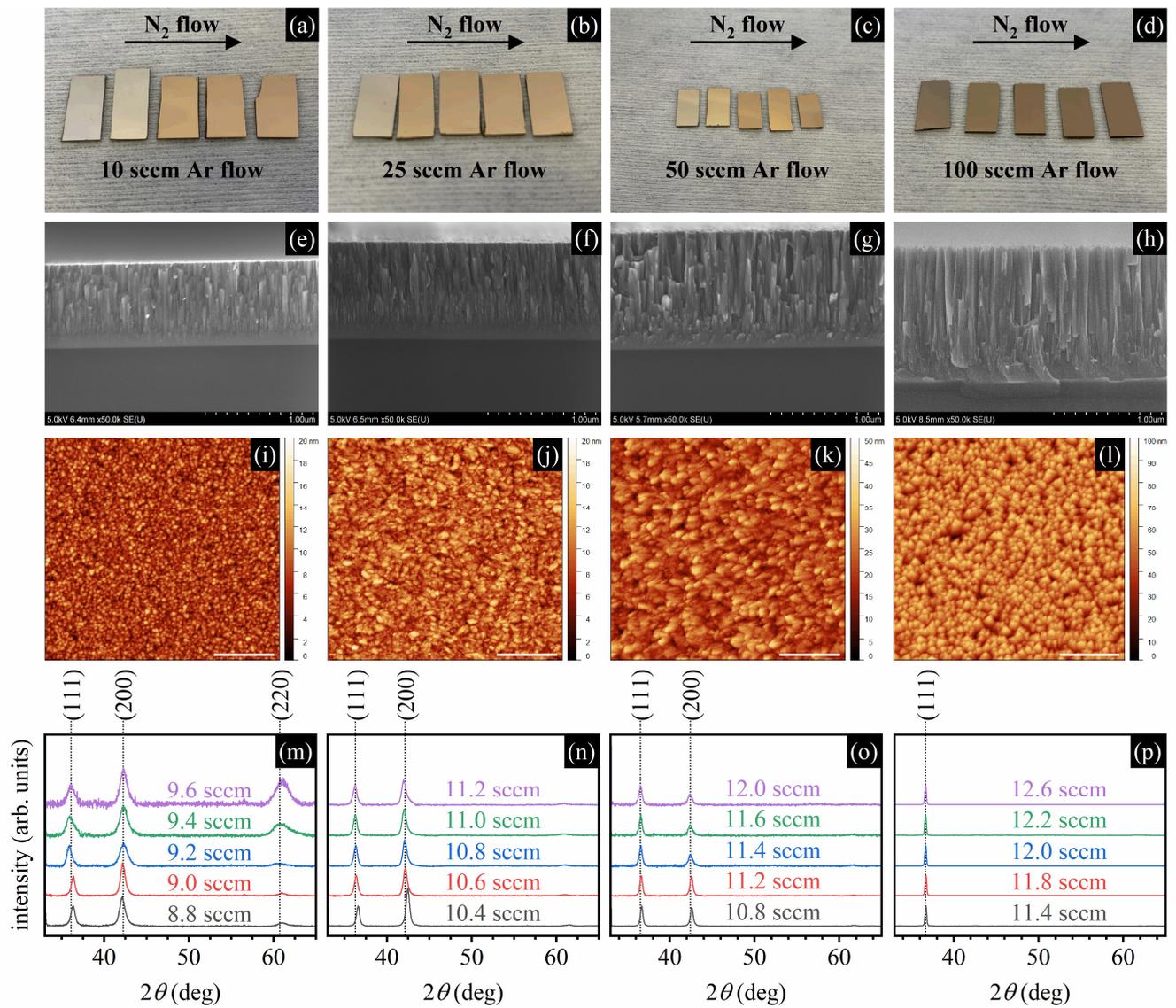

**Figure 1.** (a)-(d) Pictures of four series of TiN samples, five samples each, deposited with different $N_2$ flows (within the series) and with different Ar flows (between the series); (e)-(h) SEM pictures of the most stoichiometric samples from the series; (i)-(l) AFM pictures of the surface of the most stoichiometric samples from the series (the length of the scale bar equals 500 nm); (m)-(p) XRD patterns for different $N_2$ (within the plots) and Ar flows (between the plots).

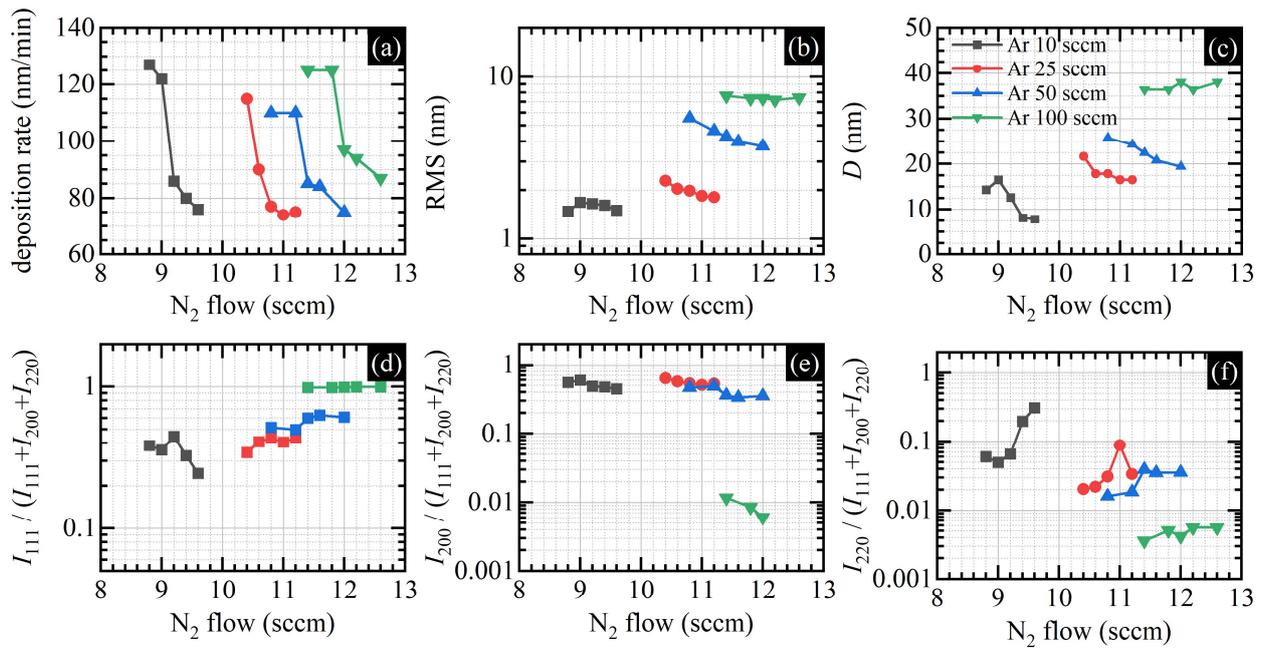

**Figure 2.** (a) Deposition rate; (b) RMS; (c) nanocrystallite size calculated with Scherrer equation; (d) relative intensity of the XRD (111) peak; (e) relative intensity of the XRD (200) peak; (f) relative intensity of the XRD (220) peak. All data are a function of $N_2$ flow for different Ar flow values.

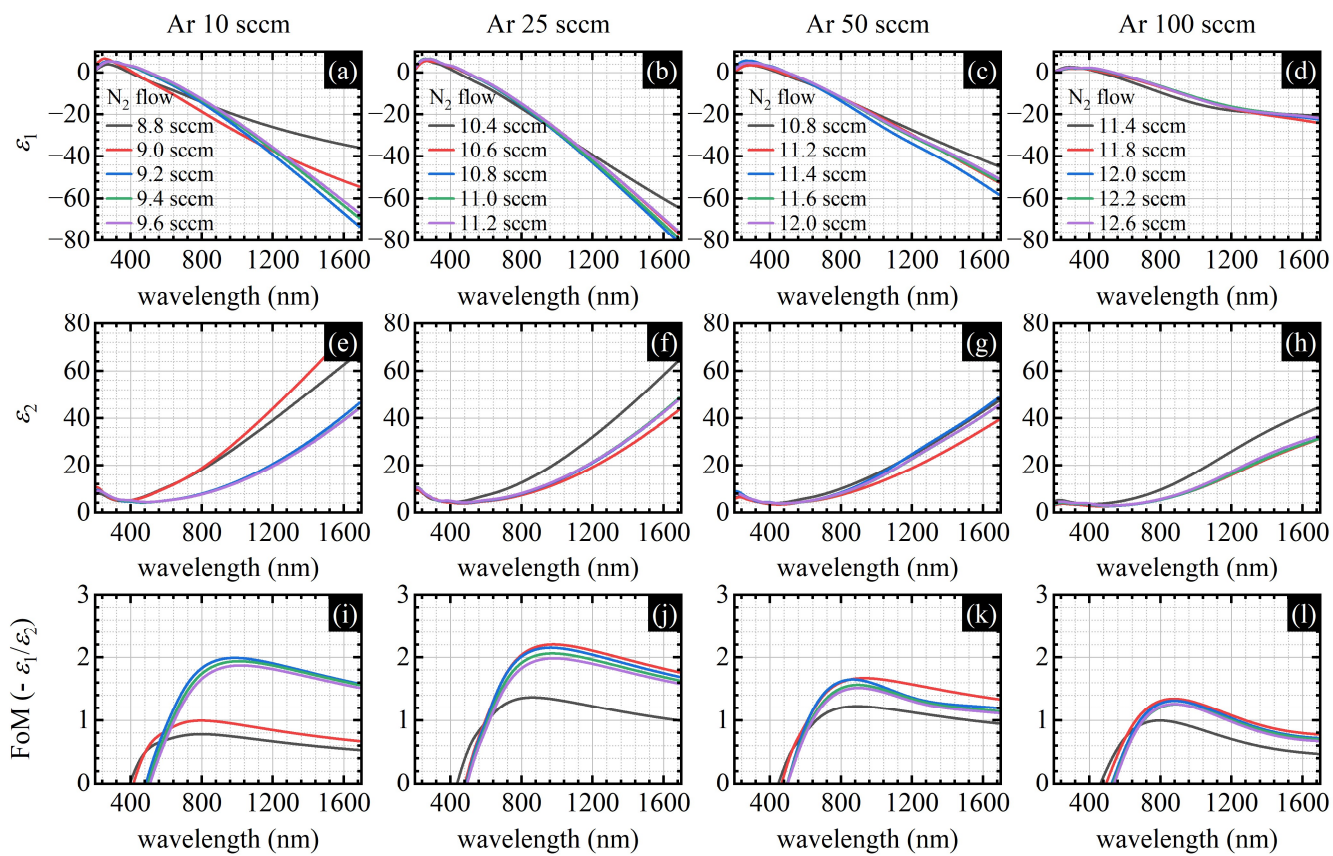

**Figure 3.** (a)-(d) Real $\varepsilon_1$ part of the dielectric function, (e)-(h) imaginary $\varepsilon_2$ part of the dielectric function, and (i)-(l) plasmonic Figure of Merit ($-\varepsilon_1/\varepsilon_2$) as a function of the wavelength for different $N_2$ (within the plots) and Ar flows (between the plots).

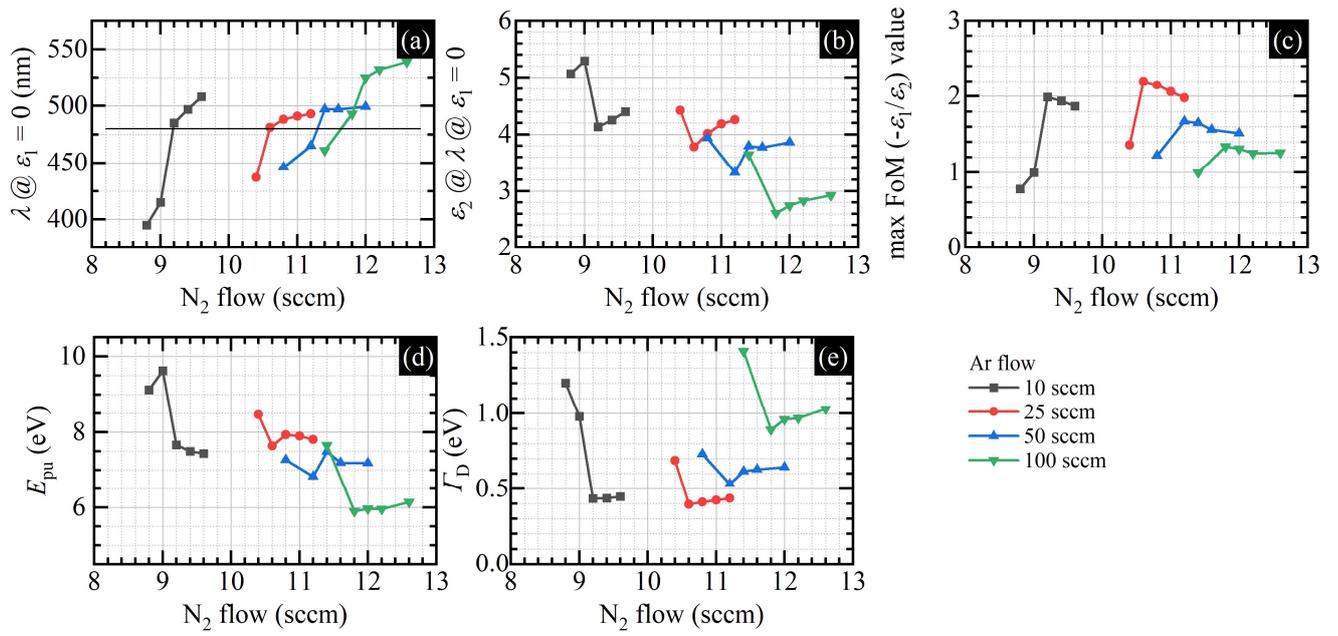

**Figure 4.** (a) Wavelength at which the real part of the dielectric function equals zero, (b) imaginary part of the dielectric function at the wavelength at which the real part of the dielectric function equals zero, (c) maximum values of the plasmonic Figure of Merit, (d) plasma energy in the Drude model, and (e) related damping coefficient of TiN films as a function of $N_2$ flow for different Ar flow values.

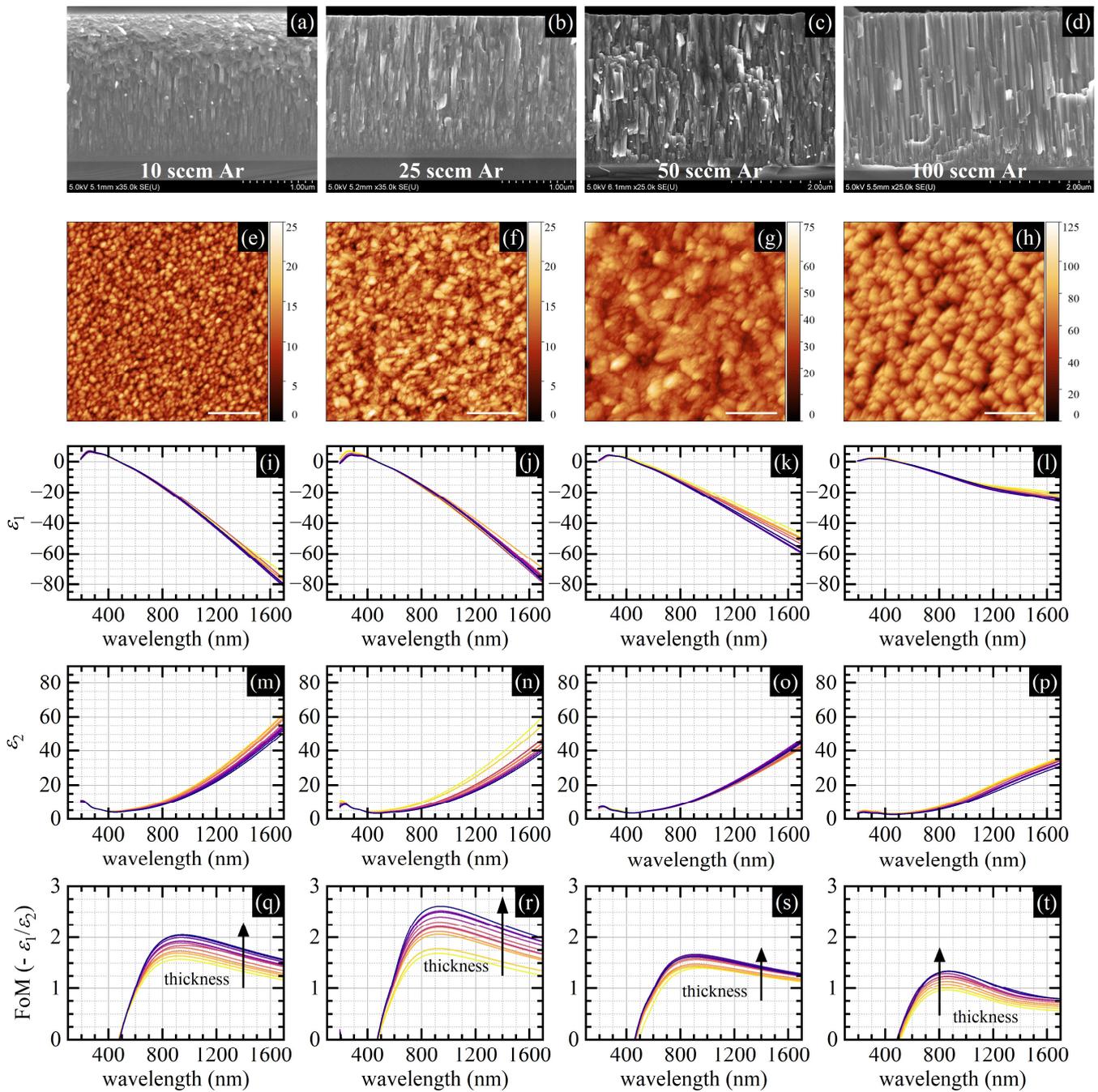

**Figure 5.** (a)-(d) SEM pictures, (e)-(h) AFM pictures (the length of the scale bar equals 500 nm), (i)-(l) real $\varepsilon_1$ part of the dielectric function, (m)-(p) imaginary $\varepsilon_2$ part of the dielectric function, and (q)-(t) the plasmonic Figure of Merit ($-\varepsilon_1/\varepsilon_2$) as a function of wavelength for different thickness values (within the plots) and Ar flows (between the plots).

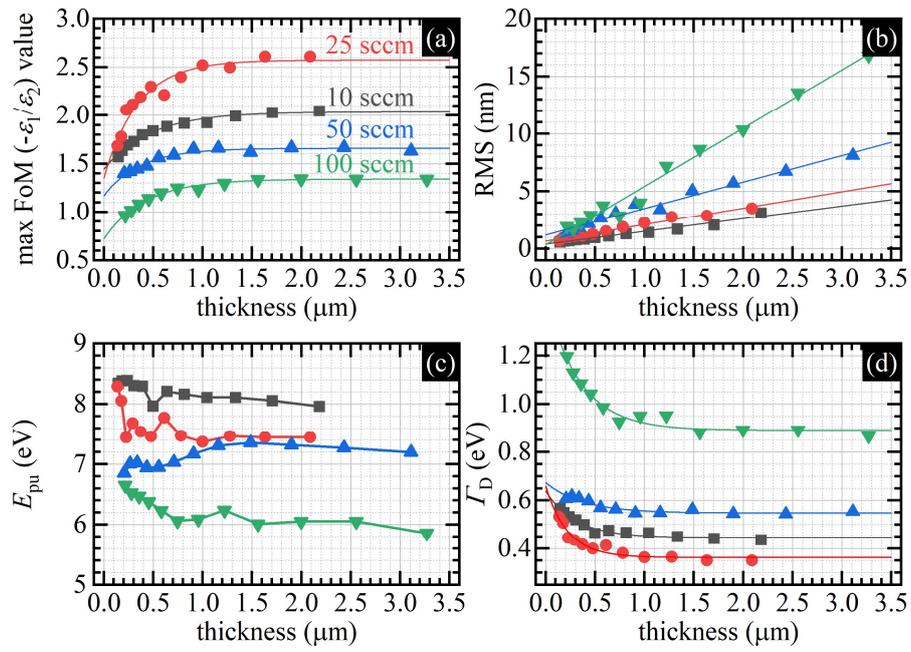

**Figure 6.** (a) Maximum values of the plasmonic Figure of Merit, (b) RMS, (c) plasma energy in the Drude model, and (d) related damping coefficient of TiN films as a function of films' thickness for different Ar flow values.

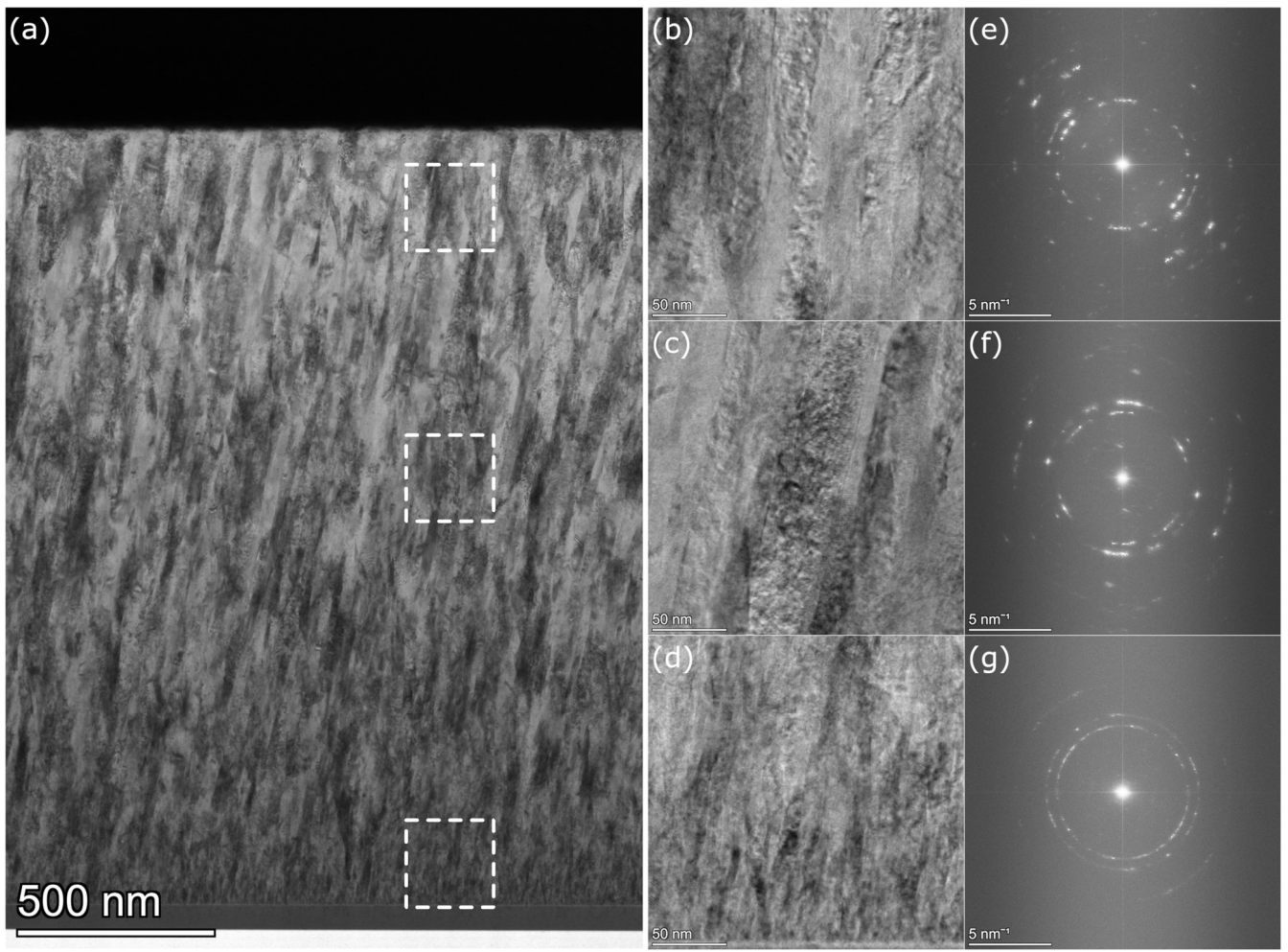

**Figure 7.** (a) STEM picture of the cross-section of the thickest sample deposited with 25 Ar flow; (b)-(d) zooms of the areas near the surface, in the middle of the layer, and near the substrate; (e)-(g) corresponding FFT diffraction patterns.

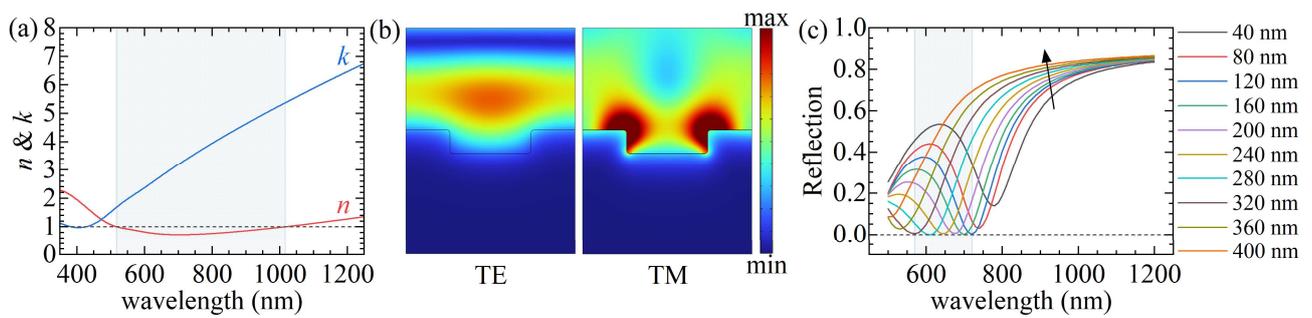

**Figure 8.** (a) Refractive index *n* and extinction coefficient *k* for sample possessing best plasmonic properties, the blue region indicates spectral range when $n < 1$, (b) electric field distribution when the polarization of the incident light is parallel (TE) and perpendicular to the nanogrooves (TM), (c) simulated reflection coefficient from plasmonic grating during normal incidence when the electric field is perpendicular to the nanogrooves, the blue region indicates the spectral range for which null reflection at particular wavelength can be observed.

# Supplementary Information: Optimization of the plasmonic properties of titanium nitride films sputtered at room temperature through microstructure and thickness control

Mateusz Nieborek[1], Cezariusz Jastrzębski[2], Tomasz Płociński[3], Piotr Wróbel[4], Aleksandra Seweryn[5], and Jarosław Judek[1,*]

[1] Institute of Microelectronics and Optoelectronics, Warsaw University of Technology, Koszykowa 75, 00-662 Warsaw, Poland.
[2] Faculty of Physics, Warsaw University of Technology, Koszykowa 75, 00-662 Warsaw, Poland.
[3] Faculty of Materials Science and Engineering, Warsaw University of Technology, Wołoska 141, 02-507, Warsaw, Poland.
[4] Faculty of Physics, University of Warsaw, Pasteura 5, 02-093 Warsaw, Poland.
[5] Institute of Physics, Polish Academy of Sciences, Aleja Lotników 32/46, 02-668 Warsaw, Poland.
[*] Corresponding author, e-mail address: jaroslaw.judek@pw.edu.pl.

**Spectroscopic ellipsometry**

In our study, the structural and optical properties of $TiN_x$ samples were analyzed as a function of $N_2$ and Ar flow as well as the deposition time. The thickness values of the analyzed in our work films range from 75 nm to 3270 nm, whereas the surface roughness ranges from 0.58 nm to 16.8 nm. Since ellipsometry is an extremely sensitive technique that can detect layers of thickness down to 0.1 Å [1] a question arises about the proper optical model to transform the experimentally obtained $\Psi$ and $\Delta$ values to the real and imaginary part of the permittivity. The simplest model assumes that the considered layer is semi-infinite and flat. In such a case, one can use the following transformation to get the pseudo-dielectric function as follows:

$$\langle \varepsilon \rangle = \sin^2 \theta \left[ 1 + \tan^2 \theta \left( \frac{1-\rho}{1+\rho} \right)^2 \right], \quad (1)$$

where $\rho = \tan \Psi e^{i\Delta}$. The first hesitation regards whether the examined films are absorbing enough that one can neglect the underneath layers, i.e., $TiO_x$ interlayer and silicon substrate with native silicon oxide layer. Whereas the answer to this question for the thickest films is positive, it turns out that for the thinnest samples, the inclusion of the physically existing layers below the examined TiN film to the optical model may contribute to the result. The second concern is related to surface roughness. This question is often neglected in the literature. Whereas for the flattest samples characterized by the RMS value below 1 nm, it may turn out that this approach is often sensible, the neglection of the surface topography for roughness value above at least a few nm, which, e.g., manifests itself during the ellipsometric measurement as a diffuse reflection, might be concerning.

To address the above-described two doubts, we performed a simple numerical experiment in which we used five optical models to transform the $\Psi$ and $\Delta$ values, which are raw measurement data from the ellipsometric measurement, to the real and imaginary parts of the permittivity. These optical models are 1) the pseudo-dielectric function approach, 2) the multi-layer model with flat surfaces, and the multi-layer model with a rough surface, where the surface roughness is modeled as an effective layer consisting in 50 % of air voids and underlying material which optical constants are calculated with the use of Bruggeman Effective Medium Approximation and characterized by the thickness value that: 3) equals the RMS value; 4) equals 1.5·RMS + 4 Å [2]; and 5) is a fitting parameter.

Figures S1 and S2 illustrate the influence of the optical model details on the extracted results. Figure S1 is related to three samples with different thicknesses deposited with 10 Ar flow, which ensures the least rough surface, whereas Figure S2 is related to three samples with different thicknesses deposited with 100 Ar flow, which leads to the most rough surface.

A comparison of those five models shows a minor influence of the underneath layers and a non-negligible influence of the roughness layer on the extracted material dielectric function. For the first series of samples, the first and second optical models give almost identical results, whereas the fifth optical model seems to stand out from the rest. In the infrared region, the real part of the permittivity takes the least negative values for the first two models and the most negative values for the fifth model. Similarly, the imaginary part of the permittivity takes the least positive values for the first two models and the most positive values for the fifth model. As a consequence, the plasmonic Figure of Merit in the case of the first two models is the highest. The inclusion of roughness leads to a slight decrease in the value of FoM and a decrease in the wavelength at which the FoM takes maximal value. For the second series of samples, particularly for the thickest sample, the inclusion of surface roughness leads to a drastic dependency of the extracted permittivity and related Figure of Merit on the selected optical model, as illustrated in Figures S2c, f, and i. But in this case, contrary to the previously analyzed series of samples, despite the real part of the permittivity being the least negative and the imaginary part of the permittivity being the least positive for the first two models, the inclusion of the surface roughness surprisingly leads to a increase in the maximal FoM value.

The proper choice of the most appropriate optical model for our samples characterized by the finite, non-negligible surface roughness is thus a difficult task. Moreover, the ellipsometric curves do not possess distinctive features that might help impose constraints on the model, and the thickness of the characterized layers prevents additional transmission measurements or usage of the interference approach [3]. To address this problem, we performed the Mean Squared Error (MSE) analysis, which results are presented in Table S1. As can be seen, the fifth optical model, i.e., the model in which the thickness of the additional, most on-top effective layer simulating surface roughness is a variable in the fitting procedure, seems to give the lowest value, suggesting it is the most accurate. And the one we are using in this paper.

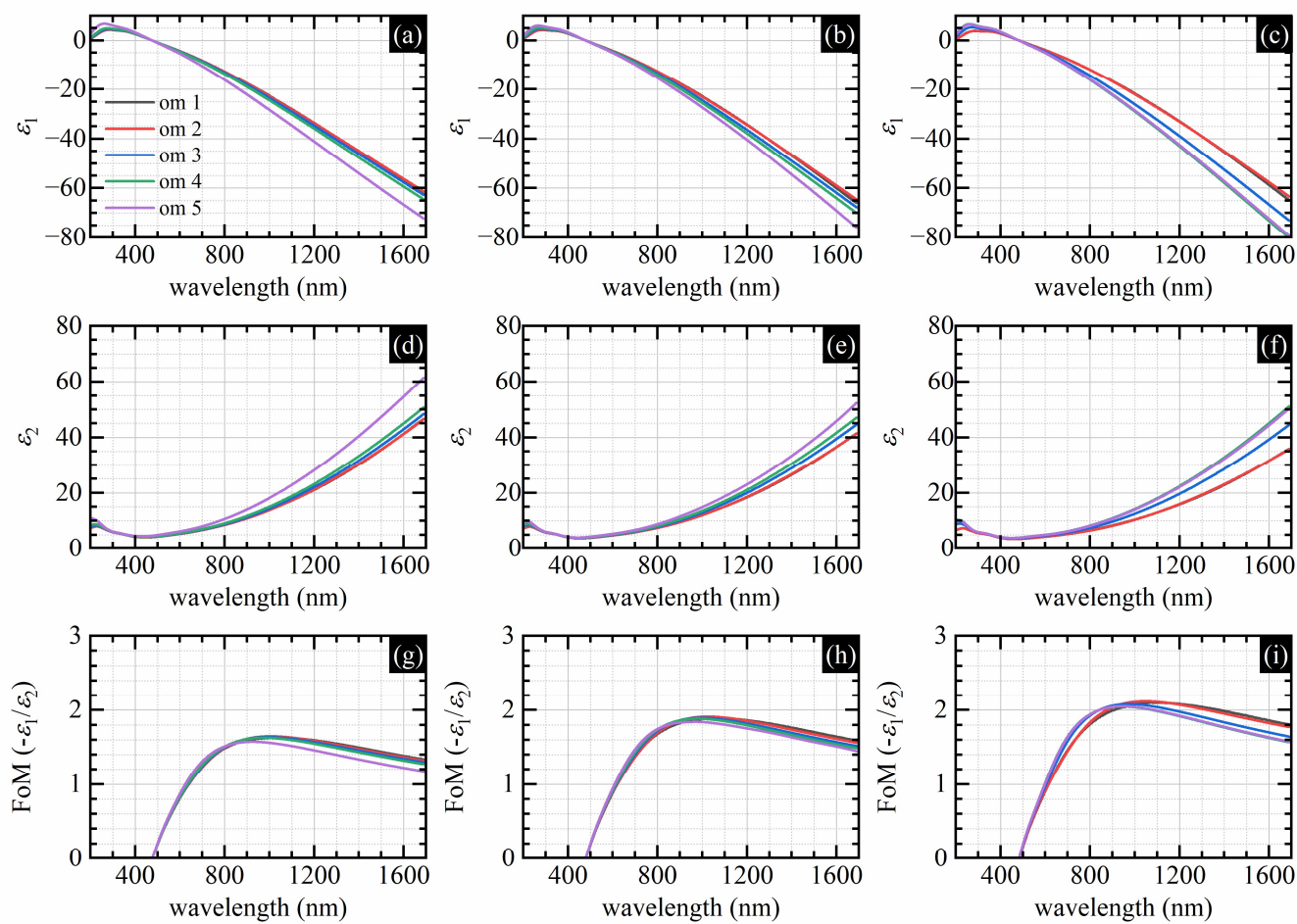

**Figure S1.** (a)-(c) Real $\varepsilon_1$ part of the dielectric function, (d)-(f) imaginary $\varepsilon_2$ part of the dielectric function, and (g)-(i) plasmonic Figure of Merit ($-\varepsilon_1/\varepsilon_2$) as a function of the wavelength for three stoichiometric samples obtained for 10 sccm Ar flow with a thickness of 145 nm, 498 nm, and 2180 nm. Different curves represent five optical models ("om") to transform the experimentally obtained raw $\Psi$ and $\Delta$ data.

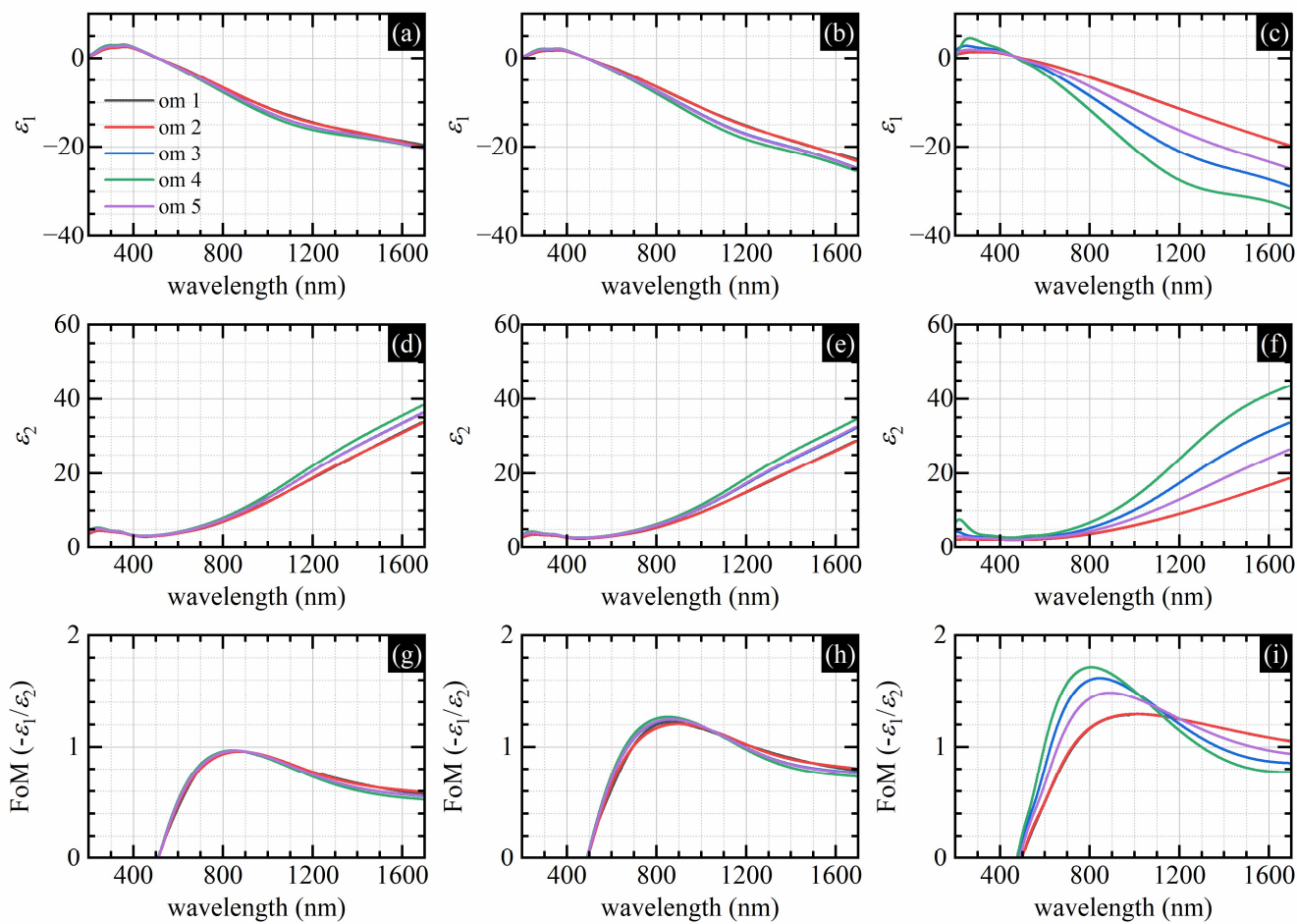

**Figure S2.** (a)-(c) Real $\varepsilon_1$ part of the dielectric function, (d)-(f) imaginary $\varepsilon_2$ part of the dielectric function, and (g)-(i) plasmonic Figure of Merit ($-\varepsilon_1/\varepsilon_2$) as a function of the wavelength for three stoichiometric samples deposited with 100 sccm Ar flow with a thickness of 218 nm, 498 nm, and 3270 nm. Different curves represent five optical models ("om") to transform the experimentally obtained raw $\Psi$ and $\Delta$ data.

| sample | MSE values for different optical models | | | |
|---|---|---|---|---|
| | #2 | #3 | #4 | #5 |
| Ar 10 sccm, thickness 145 nm, roughness 0.58 nm | 1.16 | 1.09 | 0.99 | 0.73 |
| Ar 10 sccm, thickness 498 nm, roughness 0.96 nm | 1.57 | 1.36 | 1.20 | 0.82 |
| Ar 10 sccm, thickness 2180 nm, roughness 3.10 nm | 1.93 | 1.20 | 1.01 | 1.00 |
| Ar 100 sccm, thickness 218 nm, roughness 1.96 nm | 3.59 | 3.56 | 3.63 | 3.59 |
| Ar 100 sccm, thickness 747 nm, roughness 2.87 nm | 3.87 | 3.80 | 3.78 | 3.79 |
| Ar 100 sccm, thickness 3270 nm, roughness 16.8 nm | 5.47 | 5.54 | 7.12 | 5.15 |

**Table S1.** Comparison of Mean Squared Error of fitting roughness-dependent ellipsometric models.